\documentclass[%
reprint,
nofootinbib,
 amsmath,amssymb,
 aps,
]{revtex4-2}

\usepackage{graphicx}
\usepackage{dcolumn}
\usepackage{bm}
\usepackage{color}
\usepackage{ulem}

\usepackage{subfigure}

\begin{document}

\preprint{APS/123-QED}

\title{
Thick Lunar Crust Amplifies Deci-Hertz Gravitational-Wave Signal
}

\author{Lei Zhang$^{1}$}
  \thanks{The first two authors contributed equally to this work.}
\author{Han Yan$^{2,3}$}
  \thanks{The first two authors contributed equally to this work.}
\author{Xian Chen$^{2,3}$}
  \email{Contact author: xian.chen@pku.edu.cn}
\author{Jinhai Zhang$^{1}$} 
  \email{Contact author: zjh@mail.iggcas.ac.cn}

\affiliation{$^{1}$Institute of Geology and Geophysics, Chinese Academy of Sciences, Beijing 100029, China}

\affiliation{$^{2}$Department of Astronomy, School of Physics, Peking University, 100871 Beijing, China}

\affiliation{$^{3}$Kavli Institute for Astronomy and Astrophysics, Peking University, 100871 Beijing, China}

\date{\today}

\begin{abstract}
	Gravitational waves (GWs) in the $0.01\sim1$ Hz band encode unique signatures
of the early universe and merging compact objects, but they are beyond the
reach of existing observatories.  Theoretical models suggest that the Moon
could act as a resonant detector, but the unknown influence of its rugged surface and heterogeneous interior poses a challenge to the accurate modeling of its response.  Here, we
address this long-standing uncertainty by constructing the first
high-resolution, two-dimensional model of the lunar GW response, more realistic 
than previous ones. We
achieve this by combining high-fidelity spectral-element
simulations with
the analytical power of normal-mode perturbation theory, thereby resolving
topographical effects down to 2 km grid spacing while maintaining the capacity
to discern global free-oscillation patterns.
This dual-methodology approach not only recovers the expected predominant quadrupole 
($l=2$) oscillation mode, but also exposes a systematic signal amplification in thick-crust
regions.
This enhancement is traced by our normal-mode analysis to a mode-coupling process, in which the original quadrupolar oscillation induced by the passing GW distributes energy into a series of higher-order modes, the hybridized eigenmodes of a laterally heterogeneous Moon.
In certain narrow frequency ranges, we observe up to tenfold amplification spanning into the deci-hertz band, highlighting
the power of numerical simulations in resolving these structurally fine-tuned features for designing future detectors.
Our work establishes the Moon as a resonant GW detector albeit its complex topographical structures, and the resulting
	amplification maps provide quantitative guide for the optimal landing site selection.
\end{abstract}

\maketitle


\textit{Introduction.}---Gravitational waves (GWs) have been observed in the
hundred-hertz band by ground-based interferometers \cite{2016PhRvL.116f1102A,
1992Sci...256..325A, 2015CQGra..32b4001A, 2019NatAs...3...35K} and in the
nano-hertz band by pulsar timing arrays \cite{2023RAA....23g5024X,
2023A&A...678A..50E, 2023ApJ...951L...9A, 2023ApJ...951L...6R}. In between lies
the mid band ($0.01\sim1$ Hz) where many intriguing sources exist, including
the seeds of supermassive black holes, inspiraling binary stellar black holes
or neutron stars, supernovae, as well as quantum fluctuations in the early
universe \cite{2020CQGra..37u5011A,2024arXiv240207571C}.  The detection of
these sources defines the core science cases for future space-based GW
detectors like LISA, TianQin, Taiji \cite{2024arXiv240207571C,
2016CQGra..33c5010L, 2021PTEP.2021eA108L}, and other projects \cite{2011CQGra..28i4011K,2022IJMPD..3150039N}.

An alternative proposal of detecting mid-band GWs is to find a celestial body
which can act as a giant resonant detector \cite{1960PhRv..117..306W,
1968PhT....21d..34W,1969ApJ...156..529D}. In this spirit, the Moon offers a
unique platform due to its quietness compared to the Earth
\cite{2009JGRE..11412003L}. Assuming spherical symmetry,  earlier analytical
works established that only the Moon's quadrupole ($l=2$) mode resonates with
GW \cite{1983NCimC...6...49B,1996CQGra..13.2865B}, and the resonant peaks lie
in the mid band \cite{1983NCimC...6...49B, 2019PhRvD.100d4048M,
PhysRevD.109.064092,2025PhRvD.111d4061M}. These one‑dimension models also revealed the
importance of lunar radial structure in determining the resonant frequencies
\cite{2024PhRvD.110f4025B,2025PhRvD.111d4061M}. 

In reality, the Moon is {\it not} a spherically symmetric body but has strong variations in both topography and crustal thickness, since altimetric surveys indicate tens of kilometers in topographic relief \cite{2010GeoRL..3718204S,2013Sci...339..671W} and lunar
gravity measurements reveal strong lateral variations in crustal thickness \cite{huang2022bombardment,
wieczorek1998potential}. Many studies based on
normal-mode perturbation theory \cite{1980GeoJ...61..261W,
1999tgs..book.....D} show that strong lateral velocity variations would severely impact the resonance of seismic wave propagation within the Earth, especially for the normal mode coupling and energy
redistribution \cite{1978GeoJ...53..335W,1983JGR....8810285M,2001GeoJI.146..833D}. However, the effect of such heterogeneity on the GW response of the Earth and the Moon remains unknown. 
If not
properly understood, it may mislead our conclusion for lunar GW detection \cite{Chen2026-be}. This
issue is becoming even more pressing for the proposed projects like Lunar Gravitational-Wave Antenna (LGWA) \cite{2021ApJ...910....1H,
2023SSRv..219...67B}, Laser Interferometer Lunar Antenna (LILA) \cite{2025arXiv250811631J}, and other lunar GW projects \cite{2009AdSpR..43..167P,2021CQGra..38l5008A,2023SCPMA..6609513L}.

Using conventional normal‑mode perturbation analysis \cite{1980GeoJ...61..261W,
1999tgs..book.....D} to address this issue presents distinct advantages and
limitations. A major advantage arises from the simplicity of the GW source
term: unlike earthquakes or moonquakes, which excite a broad spectrum of
seismic modes, GWs interact almost exclusively with the Moon’s quadrupole
($l=2$) modes, drastically simplifying the analytical description. However,
this approach becomes computationally demanding when accounting for the full
mode‑mixing induced by the lateral heterogeneity, and it struggles to provide a
fine‑scale description of the resulting surface displacement field.  To
complement the perturbation analysis, in the past few years we have developed
high‑fidelity numerical simulations based on the high-order finite-element method
(FEM), specifically the spectral‑element method (SEM)
\cite{2025PhRvD.111f3014Z,2026PhRvD.113b3031Z},
which enables precise modeling of wave propagation across realistic lunar
topography \cite{komatitsch2005spectral,
zhang2020procedure,2026PhRvD.113b3031Z}. Although computationally intensive,
the SEM provides a direct numerical testbed for verifying key analytical
predictions, particularly in regions of strong topographic and crustal
variations. However, three‑dimensional (3D) simulations remain computationally prohibitive for
long-duration GW signals. Therefore, it is essential to develop a novel
calculation framework that combines the high resolution of the FEM and the
generality of the normal-mode analysis
to reveal both the fine details of the lunar response and the underlying physical mechanism.

For the above reason, we present the first structurally realistic model of the lunar GW response: high-precision two‑dimensional (2D) SEM simulations employed to resolve the effects of rugged topography and strong non-uniform crustal structures, while a 3D mode-coupling approach characterizes the global low-order response. Our results show that the Moon’s
crustal asymmetry acts as a potent amplifier, efficiently redistributing energy
from the GW‑excited quadrupole modes into a rich spectrum of higher‑order
spheroidal modes. This mode‑mixing process not only enhances signal
detectability in specific frequency bands but also indicates the potential of GW‑induced seismic responses to constrain the Moon’s 3D internal structure.

\begin{figure*}[htpb]
    \centering
    \includegraphics[width=18cm]{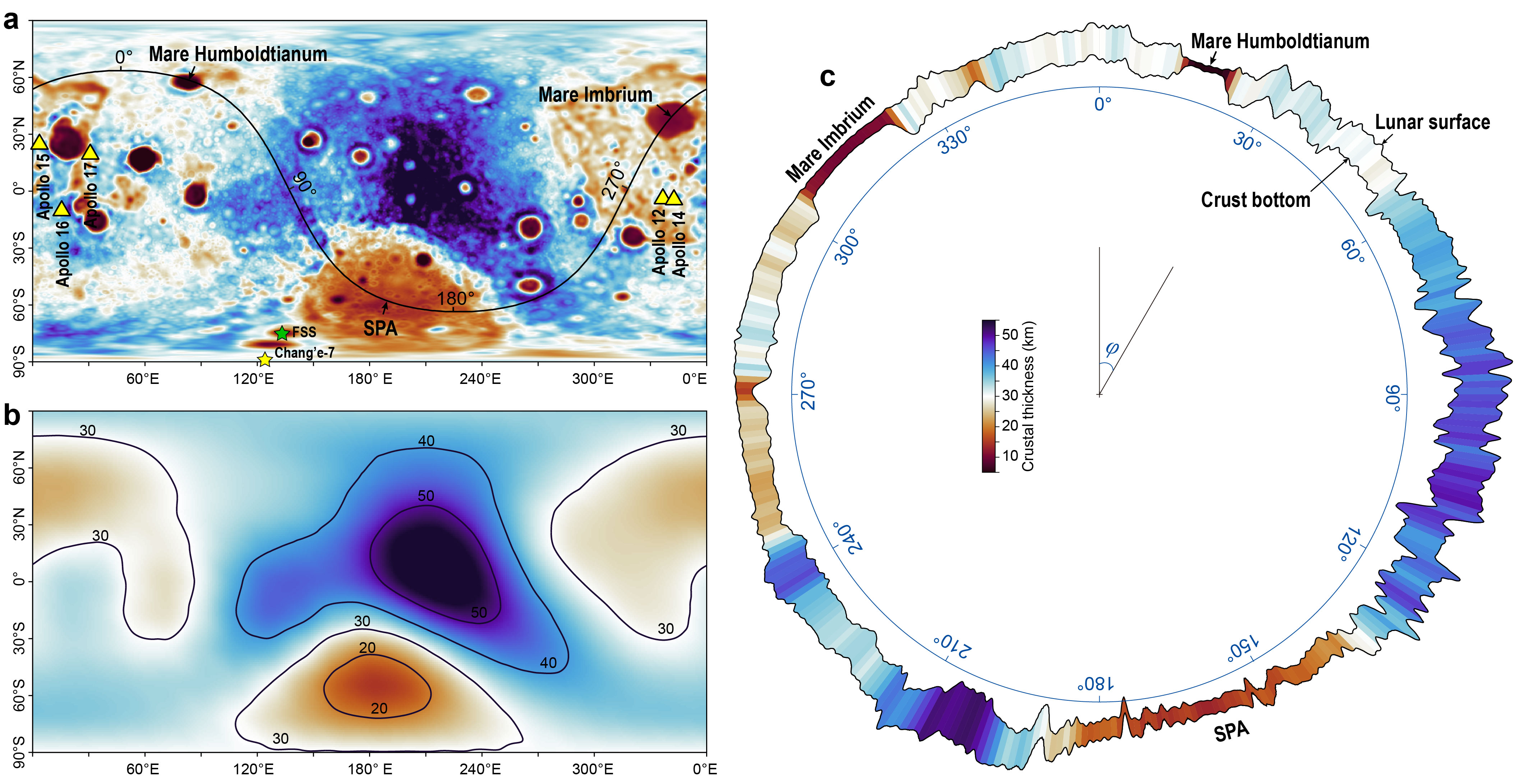}
    \caption{Lunar model with topography and crustal-thickness variations. (a) Lunar crustal thickness model. The yellow triangles mark the locations of Apollo seismographs. The black line indicates the great circle passing through Mare Humboldtianum, Mare Imbrium, and the SPA basin. The potential landing sites of Chang’e-7 and FFS are marked as stars around the south pole. (b) Crustal thickness model truncated at $l=4$. Solid lines represent crustal thickness contours in km. (c) The surface topography and crustal thickness along the great circle shown in (a). The outer black curve is the surface topography, exaggerated by a factor of 30 for visibility, and the inner black curve marks the crust bottom, whose radial position is drawn by exaggerating the crustal thickness by a factor of 5 inward from the surface, both relative to the mean lunar radius. The color fill encodes the local crustal thickness, highlighting its pronounced lateral variations. }
    \label{fig:fig1}
\end{figure*}

\begin{figure*}[ht]
    \centering
    \includegraphics[width=18cm]{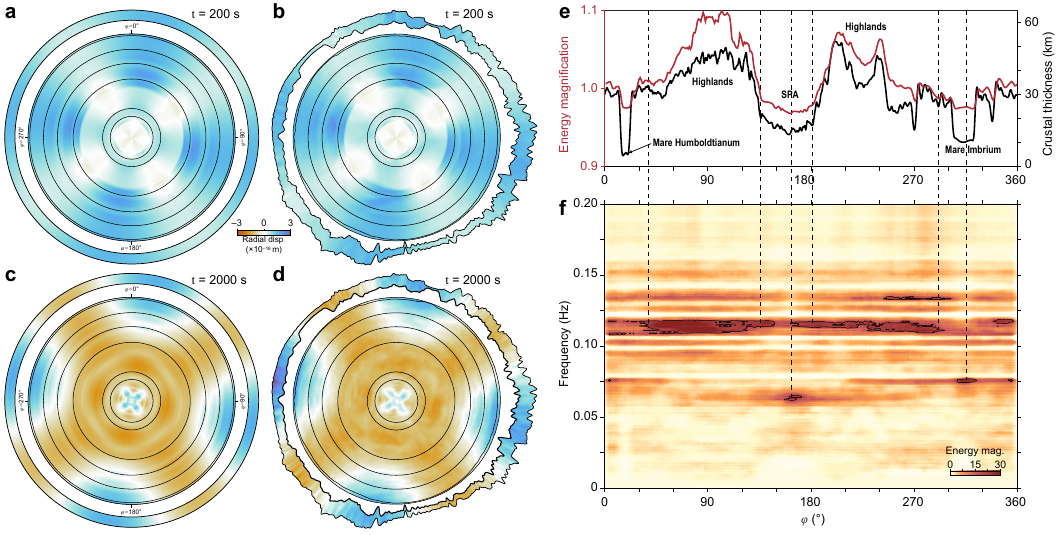}
    \caption{Comparison between layered and heterogeneous models and the simulated location-frequency distribution of amplification ratios. (a) The snapshot of lunar response to GWs for a spherically layered model at t = 200 s. (b) The same as (a) but for the model with varying topography and crustal thickness shown in Figure 1c. In a and b, the topography and crustal thickness are magnified by 30 and 5 times in the outside annulus, respectively. (c) The same as (a) but at t = 2000 s. (d) The same as (b) but at t = 2000 s. More snapshots can be found in \cite{webs}. (e) The averaged amplification ratio $M_E$ (red line), integrated up to 0.2~Hz. The black line is the crustal thickness for reference. (f) The amplification ratio $m_E (f)$ in the frequency band up to 0.2~Hz, where the black contour lines indicate the amplification ratio of 20.  }
    \label{fig:fig2}
\end{figure*}

\textit{Numerical simulation of lunar response to GWs.}---We first perform 2D
numerical simulations to quantify how lunar topography and crustal structure
modulate the GW‑induced surface displacement and velocity fields. Details of modeling and parameter setting can
be found in Supplemental Material \cite{supplemental}. Our lunar
model integrates topographic and crustal-thickness data from the LOLA
\cite{2010GeoRL..3718204S} and GRAIL \cite{2013Sci...339..671W} missions,
incorporating major structural features such as the South Pole–Aitken (SPA)
basin, Mare Imbrium, and Mare Humboldtianum (Figure~\ref{fig:fig1}). To reduce the extremely heavy computational cost, we
extract a typical 2D profile along a selected great circle of the real 3D model. Dyson-type force density is applied to model
the GW sources \cite{1969ApJ...156..529D}. To accurately capture short-wavelength 
scattering in the deci-hertz regime, this simulation utilizes a high-resolution 
topographic model with a grid spacing of approximately 2.0 km. Our model achieves a maximum
resolved frequency of 0.36 Hz with a time step constrained to $5\times 10^{-3}$ s, ensuring
both numerical stability and computational tractability. Then we use the SEM
\cite{2025PhRvD.111f3014Z} to simulate the excitation of seismic activity by
GWs, employing a broadband source time function centered near $0.1$ Hz. We focus on the response in the
frequency band up to $0.2$~Hz. Figure~\ref{fig:fig2}(a--d) shows
that the lateral heterogeneity of structures has significant impact on the
lunar response to GWs. Compared with the results obtained by spherically
layered models \cite{PhysRevD.109.064092, 2025PhRvD.111f3014Z}, the wavefronts
in the realistic model are evidently distorted, especially in Mare Imbrium, Mare
Humboldtianum, and the SPA basin. 

To quantify the impact of lunar topography and crustal thickness variation, we define an
amplification ratio $M_E$ as the ratio of local displacement energy ($|\bm{\vec{s}}(\vec{r}, \omega)|^2$) \footnote{Note that $\omega=2\pi f$. The frequency units ``mHz'' or ``Hz'' in this work refer to
$f$ rather than the angular frequency $\omega$.}
with or without lateral heterogeneity (marked with ``P'' or ``NP''), integrated over frequency: 
\begin{eqnarray}
\label{eq:magnification}
M_E(\vec{r}) & = & \frac{ \int_{\omega_{\text{min}}}^{\omega_{\text{max}}} \langle |\bm{\vec{s}}^{\text{P}}(\vec{r}, \omega)|^2 \rangle_{\text{pol}} \mathrm{d}\omega }{ \int_{\omega_{\text{min}}}^{\omega_{\text{max}}} \langle |\bm{\vec{s}}^{\text{NP}}(\vec{r}, \omega)|^2 \rangle_{\text{pol}} \mathrm{d}\omega } ~.
\end{eqnarray}
Here, $\langle \rangle_{\text{pol}}$ denotes an average over all polarization, in order to
separate the amplification due to lunar structure from polarization-induced variations in the GW amplitude. More details of this equation are discussed in the Supplemental Material \cite{supplemental}.

Figure~\ref{fig:fig2}e presents the distribution of $M_E$ along the typical great circle (integrated up to 0.2~Hz). The spatial characteristics generally show strong correlations
with both topography and crustal thickness. Notably, thick-crust regions
(e.g., highlands) show an average signal amplification (red curve in Figure~\ref{fig:fig2}e). However, in some places
such as Mare Imbrium, where the topography has almost no relief, the
amplification ratios still vary dramatically with the varying crustal
thickness. This key observation demonstrates that crustal thickness rather than
topography is the dominant factor of governing the amplification effect. Hence, we
should pay more attention to the crustal thickness when evaluating lunar local
response to GWs in future works.

To find the optimal frequency band for GW search, we plot the
frequency-dependent amplification ratio $m_E (f)$, defined as
\begin{eqnarray}
\label{eq:freq-magnification}
m_E(\vec{r},f) & = & \frac{ \langle |\bm{\vec{s}}^{\text{P}}(\vec{r}, 2\pi f)|^2 \rangle_{\text{pol}}  }{  \langle |\bm{\vec{s}}^{\text{NP}}(\vec{r}, 2\pi f)|^2 \rangle_{\text{pol}}  } ~,
\end{eqnarray}
in Figure~\ref{fig:fig2}f, where we have applied a median-filter smoothing to
reveal fundamental patterns. The plot shows that the structurally induced
signal enhancement is a broadband phenomenon. In the lower frequency regime
($\le 0.04$~Hz), the amplification exhibits rich spectral peaks with narrow
bandwidths [also see Figs.~(A3) and (A4) \cite{supplemental}]. These peak
frequencies align closely with the resonant frequencies of the lunar response
function \cite{2025PhRvD.111f3014Z}, suggesting a direct link to the normal
modes of the Moon. Crucially, near the optimal sensitivity band of proposed
lunar antennas (around $0.1$ Hz), this localized energy magnification
$m_{E}(f)$ not only persists but further intensifies. The broadband spatial
variations in this deci-hertz regime remain tightly correlated with the
thick-crust topography. For example, on the highlands surrounding the SPA basin
(between $\varphi=70^{\circ}$ and $120^{\circ}$), $m_E (f)$ can reach tenfold,
with peak amplifications often exceeding those observed at lower frequencies.


\textit{Normal mode perturbation method.}---The observed close correspondence
between crust thickness and GW signal amplification in our numerical
simulations can be understood in the theoretical framework of normal-mode
perturbation and mixing. In the ideal case where the Moon is spherically symmetric, non-rotating, elastic, 
and isotropic (SNREI),
its free oscillation can be 
described by a series of orthonormal normal modes $|k\rangle \equiv |\sigma, n, l, m\rangle$ which are eigenfunctions of the secular equation 
\begin{eqnarray}
\mathcal{V} |k\rangle = \omega_k^2 \mathcal{T}|k\rangle
\end{eqnarray}
\cite{1999tgs..book.....D}.
Here, $\sigma = S/T$ indicates the mode type (Spheroidal or Toroidal), $n$ is the radial quantum number, and $(l,m)$ are the angular momentum quantum number and magnetic quantum number. $\mathcal{V}$ and $\mathcal{T}$ are potential and kinetic operators, respectively.
The presence of surface topography and crustal variation perturbs the operators by $\delta \mathcal{V}$ and $\delta \mathcal{T}$,
restructuring the eigenmodes. Given that the perturbation is small, the new modes can be approximated by a linear combination of the
basic SNREI modes, so that
\begin{eqnarray}
\label{eq:hybridization}
|i\rangle & = & \sum_k C_{ki} |k\rangle~.
\end{eqnarray}
For this reason, the new modes are also known as the ``hybridized modes''.
The transformation matrix $\mathbf{C}$ satisfies the perturbed equation
\begin{eqnarray}
\label{eq:eigenproblem}
\sum_{k'} (\omega_k^2 \delta_{kk'} + V_{kk'}) C_{k'i} & = & \Omega_i^2 \sum_{k'} (\delta_{kk'} + T_{kk'}) C_{k'i} \label{eq:eig}
\end{eqnarray}
\cite{2001GeoJI.146..833D}, where $\Omega_i$ denotes the perturbed
eigenfrequency, $V_{kk'} = \langle k | \delta\mathcal{V} | k' \rangle$ and
$T_{kk'} = \langle k | \delta\mathcal{T} | k' \rangle$ are the elements of the
coupling matrices which can be calculated using Woodhouse kernels
\cite{1980GeoJ...61..261W}. 

GWs exert a tidal force $\bm{\vec{f}}_{\text{GW}}$ on the Moon. Owing to their quadrupole character, this force  
couples almost exclusively to the unperturbed $l=2$ modes. 
In a laterally heterogeneous Moon, however, these unperturbed SNREI modes are no longer the true eigenmodes. Instead, 
they hybridize according to Equation~(\ref{eq:hybridization}). Consequently, the energy initially in the
quadrupole modes is systematically redistributed into the hybridized mode spectrum. 
Following the standard Green-function approach \cite{1983NCimC...6...49B,2019PhRvD.100d4048M} as well as perturbation formalism introduced above, this mode mixing process is captured by the following equation,
\begin{eqnarray}
\label{eq:response}
\bm{\vec{s}}^{\text{P}}(\vec{r}, \omega) & = & \sum_i \frac{\langle \vec{r}|i\rangle \langle i | \bm{\vec{f}}_{\text{GW}} \rangle}{\Omega_i^2 - \omega^2} \nonumber \\& = &\sum_i \frac{\sum_{k}C_{ki}\bm{\vec{u}}_k(\vec{r})}{\Omega_i^2 - \omega^2} \left( \sum_{k' \in \{\sigma l=S2\}} C_{k'i}^* f_{k'} \right) ~,
\end{eqnarray}
where $\bm{\vec{s}}^{\text{P}}(\vec{r}, \omega)$ is the seismic displacement
field at location $\vec{r}$, $\bm{\vec{u}}_k(\vec{r}) = \langle \vec{r}|k\rangle $ is the unperturbed
free-oscillation pattern, and $f_{k'}$ is the projection of the GW force
onto the unperturbed mode $|k'\rangle$. Other details of this core
equation can be found in the Supplemental Material C, D and F \cite{supplemental}.

We implement this framework using the topography model derived from
NASA's GRAIL Model 2 data \cite{2013Sci...339..671W}, which is already visualized in
Fig.~\ref{fig:fig1}(a). For comparison, the one-dimension SNREI model is
adopted from our previous work with slight modifications \cite{webs}. To maintain computational tractability while preserving
key large-scale asymmetries, notably the farside–nearside dichotomy, we truncate
the spherical-harmonic expansion of the Moon at $l_{\rm max}=4$, yielding the
approximate lunar model shown in Fig.~\ref{fig:fig1}(b). In the hybridized mode
calculation, we allow free-oscillation modes up to  $l=6$ and $n=100$,
reflecting both the selection rule and numerical feasibility. Toroidal modes
are excluded because GWs couple predominantly to spheroidal modes, and toroidal
eigenfunctions exhibit lower accuracy in standard solvers such as MINEOS
\cite{mineos2011}. While computing resource limits this perturbation analysis to a relatively narrow frequency band under $\sim 0.02~$Hz \footnote{Mode-perturbation approach calculations are done within two hours on a personal desktop computer without a dedicated GPU, showing its advantage on narrow-band, low-frequency global (full-Moon) analysis.}, the physical mechanism of energy redistribution validated here provides the fundamental explanation for the broadband amplification observed in our deci-hertz numerical results.

\begin{figure}
\includegraphics[width=0.98\linewidth]{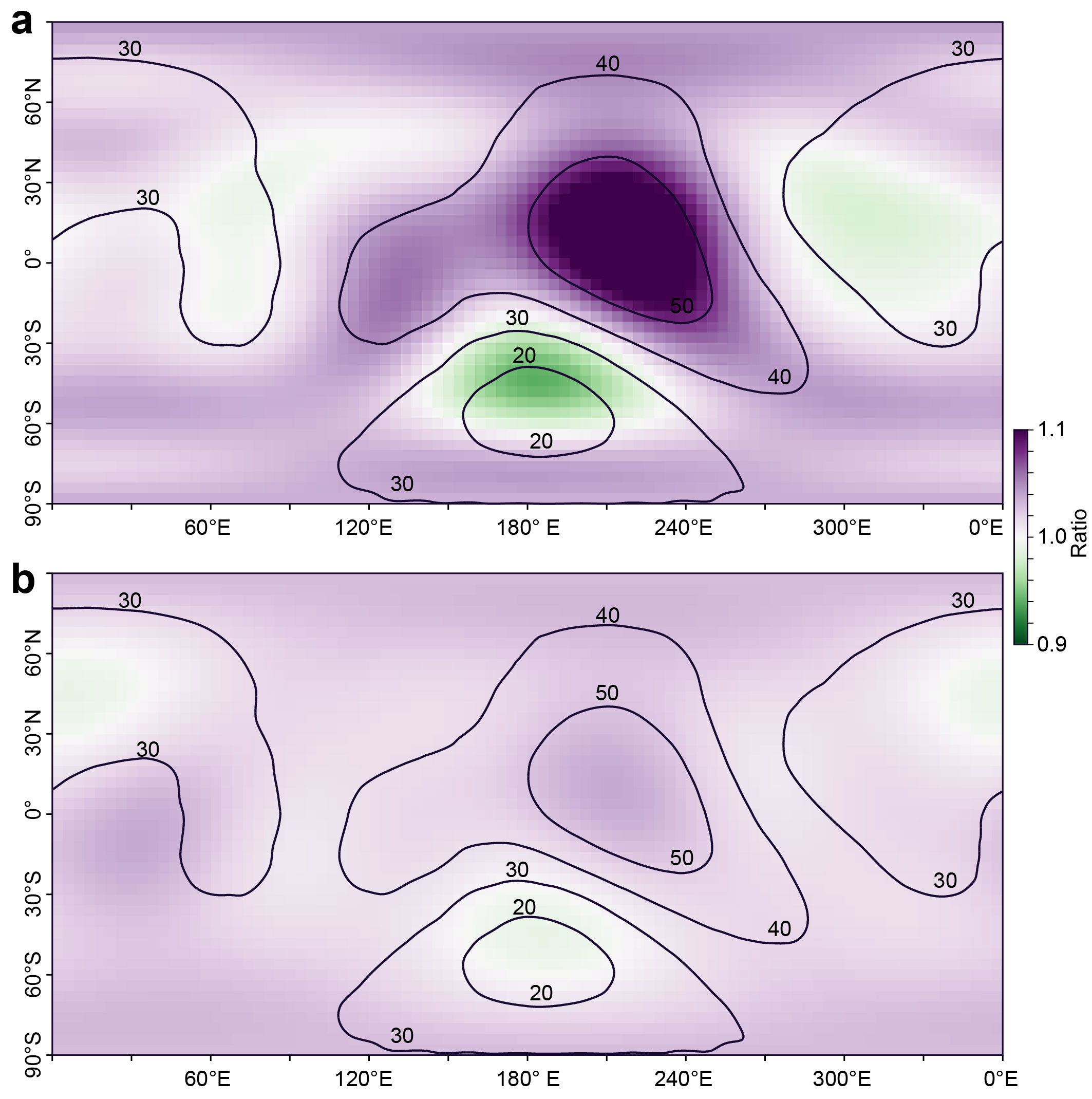}
\caption{Amplification according to the normal-mode perturbation theory. (a) Total energy magnification $M_E$, integrated between 0.01~Hz and 0.012~Hz\footnote{Same as the frequency range in Fig.~(A3) \cite{supplemental}.} and averaged over polarization. Here, all the free-oscillation modes up to $l=6$ are included. (b) Contribution
from the $l=2$ free-oscillation mode alone. For comparison, the crustal thickness contours from Fig. 1(b) are overlaid.}
\label{fig:fig3}
\end{figure}

Fig.~\ref{fig:fig3}(a) shows the band-integrated total magnification
$M_E(\vec{r})$ across the lunar surface.  This map reveals a robust and
systematic pattern, with values deviating from unity by approximately
$10\%$, consistent with the crustal-thickness dependence observed in our 2D numerical simulations.  This spatial pattern exhibits a clear correlation with the underlying
crustal structure shown in Fig.~\ref{fig:fig1}(b): regions of thin crust, such
as the nearside maria, consistently exhibit de‑amplification ($M_E<1$), whereas
the thick‑crust terrains of the farside highlands show pronounced
amplification. This correlation confirms the role of crustal thickness in
modulating the Moon’s global seismic response to GWs. It also points to a
direct link between lateral heterogeneity and surface observables.

To more clearly trace the cause of the amplification, Fig.~\ref{fig:fig3}(b)
isolates the contribution to $M_E$ from the fundamental quadrupole mode
$\bm{\vec{u}}_{l=2}$ alone.  The resulting map shows values near unity, with
much smaller variation ($\sim4\%$) and weaker correlation to crustal
thickness. This result implies that the amplification pattern seen in
Fig.~\ref{fig:fig3}(a) is not primarily due to modulation of the $l=2$ mode,
but rather emerges from the collective response of higher-order modes ($l>2$)
excited through mode mixing.

\textit{Discussions and implications.}---In this Letter, we have elucidated how
the rugged surface and heterogeneous crust of the Moon can modulate the surface
seismic signals triggered by GWs. We achieve this by developing
a calculation framework which combines the advantages of
two complementary approaches: a perturbation theory provides an efficient and transparent
physical picture of global mode mixing, whereas the SEM delivers the high spatial
resolution required to capture complex local wave-field distortions at the
expense of greater computational cost.
 
Our SEM simulation not only recovers the GW-induced quadrupole mode of the
lunar free oscillation but also reveals a close link of the seismic amplitude
to the crustal thickness. On average, the seismic signal can be amplified by
$10\%$ in thick-crust regions. In some narrow frequency ranges, the
amplification can even exceed one order of magnitude. Crucially, our
high-frequency simulations confirm that this strong amplification persists
around $0.1$ Hz, agreeing well with the optimal sensitivity window of
proposed lunar seismometer arrays.  These findings establish a direct link
between crustal structure and GW detectability. Regions characterized by
thicker crust, particularly the farside highlands, consistently exhibit an
amplification of the GW signal, therefore offer better opportunities for GW
detection. However, the relationship between topography and signal
amplification is highly complex and frequency-dependent. While broadband
integrated magnification correlates positively with crustal thickness, our
high-resolution simulations reveal that in certain narrow frequency bands
(e.g., around 0.06 Hz), local topography can induce strong focusing effects probably caused by trapped waves in basins,
leading to significant signal amplification even in regions with a relatively
thin crust. This structural complexity underscores the importance of our
2D/3D modeling in real application scenarios, such as LGWA or LILA. Optimal site
selection for seismometers cannot rely solely on macro-scale
crustal thickness; it requires mission-specific, high-resolution numerical
simulations tuned to the detectors' target sensitivity bands to fully exploit
these localized, frequency-dependent amplification zones.

Our simulations also pinpoint the strongest signal amplification in a
couple of frequency bands that can be as narrow as $\sim0.1\ \mathrm{mHz}$ [Fig.~(A3) \cite{supplemental}]. Resolving such narrow features requires
long-duration, low-noise observations since a bandwidth of $0.1\ \mathrm{mHz}$
implies a minimum observational period of $T\sim 1/(0.1~{\rm
mHz})= 10^{4}\ \mathrm{s}\approx 2.8\ \mathrm{h}$. Therefore,
continuous seismic data collection spanning one day or longer is recommended.

The correlation between signal amplification and crust thickness is also
verified by our analytical normal-mode perturbation model.  The analysis further
identifies mode mixing as the origin of the enhancement: the Moon's crustal
asymmetry reshapes its eigenmodes, enabling the energy from a purely
quadrupolar perturbation, such as that imposed by a passing GW, to transfer into
an ensemble of higher-order free-oscillation modes. Such a correlated,
structured signal across the lunar surface differs fundamentally from the
stochastic, spatially incoherent background of seismic noise. By employing a
spatial correlation approach (e.g., see Ref.~\cite{2024PhRvD.110d3009Y}) that incorporates
our theoretical model of the hybridized modes, we can in principle
statistically separate the GW-induced seismic response from the spectrally and
spatially diffuse lunar seismic background.

Several caveats are worth mentioning. Although computational costs restricted
our SEM simulations to 2D, previous 2D-vs-3D comparisons
\cite{zhang2022dichotomy} support the reliability of this approach.
Furthermore, despite differing dimensionalities, the 2D SEM and fully 3D
mode-coupling analysis yield consistent, crustal-thickness-dependent
amplification patterns at low frequencies (Figs.~\ref{fig:fig2} and
\ref{fig:fig3}). This agreement suggests the 2D great-circle model captures the
dominant GW response, since the $l=2$ driving force and leading hybridized
modes project strongly onto this plane, implying that 3D SEM simulations will
largely preserve these qualitative trends. Analytically, the mode-coupling
calculation excluded toroidal modes and was truncated at a relatively low
spherical harmonic degree. Furthermore, our current model does not resolve
the km- to sub-km-scale heterogeneity or the pervasive fracturing of the lunar
megaregolith \cite{ZhangX2022,Onodera2022,Wiggins2019}.  Including them may affect
the lunar response at the
the upper end of the deci-hertz band, where the seismic wavelength becomes comparable to
these small scatterers. In this regime, multiple scattering randomizes the
phase of the GW‑induced wavefield and, as a result, prolongs the seismic signal
\cite{Latham1970,Nakamura1983}.

Future work should bridge these gaps by performing fully 3D SEM simulations and extending the mode-coupling basis to higher $l_{\max}$. Additionally, with finer topographical models and expanded computational resources, subsequent simulations could probe higher-frequency bands, where small-scale scattering and fracturing 
are expected to enhance the spatial and spectral variability of the local response. Finally, while this study focuses on polarization-averaged global correlations, individual GW events may exhibit sharper local resonances due to constructive interference; modeling these complex, event-specific interactions will be crucial for precise source parameter estimation in future observations.

Finally, we emphasize that the integration of global mode‑coupling theory with
high‑resolution SEM simulations provides the framework necessary to reach the target of calibrating
the Moon as a resonant GW antenna. Beyond enabling the detection of mid‑band GW
sources, this approach also works reciprocally. In principle, observed GW signals could be analyzed to constrain the Moon’s 3D internal structure \cite{2025arXiv250915452P}, complementing traditional seismic surveys. This interdependence hints at a promising future for lunar GW seismology as a dual‑purpose diagnostic tool.

\vspace*{0.5cm}

\section*{acknowledgments}
This work is supported by the National Natural Science Foundation of China (NSFC) Grants Nos. 42325406 and 42204178, and the National Key Research and Development Program of
China Grant No. 2024YFC2207300. Han Yan acknowledges support from the China Scholarship Council (No.~202506010256), and thanks Gran Sasso Science Institute for hospitality during the manuscript revision. The authors would like to thank many people for helpful discussions: Jan Harms for various ideas about lunar GW detection, Li Zhao and Yanbin Wang for mode-coupling theory, and Zhaorong Fu for matrix computations.

J. Zhang and X. Chen conceived and designed the research. L. Zhang performed the numerical SEM simulations and amplification analysis. H. Yan developed the theoretical mode-coupling framework and conducted the semi-analytical calculations. L. Zhang and H. Yan contributed equally to the data analysis and drafted the manuscript. All authors contributed to the interpretation of results and the revision of the paper.

The elevation data is derived from Lunar Orbiter Laser Altimeter (LOLA,
https://pds-geosciences.wustl.edu/missions/lro/lola.htm). The lunar crustal thickness model is derived from https://github.com/MarkWieczorek/ctplanet.

\providecommand{\noopsort}[1]{}\providecommand{\singleletter}[1]{#1}%


\begin{thebibliography}{55}%
\makeatletter
\providecommand \@ifxundefined [1]{%
 \@ifx{#1\undefined}
}%
\providecommand \@ifnum [1]{%
 \ifnum #1\expandafter \@firstoftwo
 \else \expandafter \@secondoftwo
 \fi
}%
\providecommand \@ifx [1]{%
 \ifx #1\expandafter \@firstoftwo
 \else \expandafter \@secondoftwo
 \fi
}%
\providecommand \natexlab [1]{#1}%
\providecommand \enquote  [1]{``#1''}%
\providecommand \bibnamefont  [1]{#1}%
\providecommand \bibfnamefont [1]{#1}%
\providecommand \citenamefont [1]{#1}%
\providecommand \href@noop [0]{\@secondoftwo}%
\providecommand \href [0]{\begingroup \@sanitize@url \@href}%
\providecommand \@href[1]{\@@startlink{#1}\@@href}%
\providecommand \@@href[1]{\endgroup#1\@@endlink}%
\providecommand \@sanitize@url [0]{\catcode `\\12\catcode `\$12\catcode `\&12\catcode `\#12\catcode `\^12\catcode `\_12\catcode `\%12\relax}%
\providecommand \@@startlink[1]{}%
\providecommand \@@endlink[0]{}%
\providecommand \url  [0]{\begingroup\@sanitize@url \@url }%
\providecommand \@url [1]{\endgroup\@href {#1}{\urlprefix }}%
\providecommand \urlprefix  [0]{URL }%
\providecommand \Eprint [0]{\href }%
\providecommand \doibase [0]{https://doi.org/}%
\providecommand \selectlanguage [0]{\@gobble}%
\providecommand \bibinfo  [0]{\@secondoftwo}%
\providecommand \bibfield  [0]{\@secondoftwo}%
\providecommand \translation [1]{[#1]}%
\providecommand \BibitemOpen [0]{}%
\providecommand \bibitemStop [0]{}%
\providecommand \bibitemNoStop [0]{.\EOS\space}%
\providecommand \EOS [0]{\spacefactor3000\relax}%
\providecommand \BibitemShut  [1]{\csname bibitem#1\endcsname}%
\let\auto@bib@innerbib\@empty
\bibitem [{\citenamefont {{Abbott}}\ \emph {et~al.}(2016)\citenamefont {{Abbott}}, \citenamefont {{Abbott}}, \citenamefont {{Abbott}} \emph {et~al.}}]{2016PhRvL.116f1102A}%
  \BibitemOpen
  \bibfield  {author} {\bibinfo {author} {\bibfnamefont {B.~P.}\ \bibnamefont {{Abbott}}}, \bibinfo {author} {\bibfnamefont {R.}~\bibnamefont {{Abbott}}}, \bibinfo {author} {\bibfnamefont {T.~D.}\ \bibnamefont {{Abbott}}}, \emph {et~al.},\ }\bibfield  {title} {\bibinfo {title} {{Observation of Gravitational Waves from a Binary Black Hole Merger}},\ }\href {https://doi.org/10.1103/PhysRevLett.116.061102} {\bibfield  {journal} {\bibinfo  {journal} {\prl}\ }\textbf {\bibinfo {volume} {116}},\ \bibinfo {eid} {061102} (\bibinfo {year} {2016})},\ \Eprint {https://arxiv.org/abs/1602.03837} {arXiv:1602.03837 [gr-qc]} \BibitemShut {NoStop}%
\bibitem [{\citenamefont {{Abramovici}}\ \emph {et~al.}(1992)\citenamefont {{Abramovici}}, \citenamefont {{Althouse}} \emph {et~al.}}]{1992Sci...256..325A}%
  \BibitemOpen
  \bibfield  {author} {\bibinfo {author} {\bibfnamefont {A.}~\bibnamefont {{Abramovici}}}, \bibinfo {author} {\bibfnamefont {W.~E.}\ \bibnamefont {{Althouse}}}, \emph {et~al.},\ }\bibfield  {title} {\bibinfo {title} {{LIGO: The Laser Interferometer Gravitational-Wave Observatory}},\ }\href {https://doi.org/10.1126/science.256.5055.325} {\bibfield  {journal} {\bibinfo  {journal} {Science}\ }\textbf {\bibinfo {volume} {256}},\ \bibinfo {pages} {325} (\bibinfo {year} {1992})}\BibitemShut {NoStop}%
\bibitem [{\citenamefont {{Acernese}}\ \emph {et~al.}(2015)\citenamefont {{Acernese}}, \citenamefont {{Agathos}} \emph {et~al.}}]{2015CQGra..32b4001A}%
  \BibitemOpen
  \bibfield  {author} {\bibinfo {author} {\bibfnamefont {F.}~\bibnamefont {{Acernese}}}, \bibinfo {author} {\bibfnamefont {M.}~\bibnamefont {{Agathos}}}, \emph {et~al.},\ }\bibfield  {title} {\bibinfo {title} {{Advanced Virgo: a second-generation interferometric gravitational wave detector}},\ }\href {https://doi.org/10.1088/0264-9381/32/2/024001} {\bibfield  {journal} {\bibinfo  {journal} {Classical and Quantum Gravity}\ }\textbf {\bibinfo {volume} {32}},\ \bibinfo {eid} {024001} (\bibinfo {year} {2015})},\ \Eprint {https://arxiv.org/abs/1408.3978} {arXiv:1408.3978 [gr-qc]} \BibitemShut {NoStop}%
\bibitem [{\citenamefont {{Kagra Collaboration}}\ \emph {et~al.}(2019)\citenamefont {{Kagra Collaboration}}, \citenamefont {{Akutsu}}, \citenamefont {{Ando}} \emph {et~al.}}]{2019NatAs...3...35K}%
  \BibitemOpen
  \bibfield  {author} {\bibinfo {author} {\bibnamefont {{Kagra Collaboration}}}, \bibinfo {author} {\bibfnamefont {T.}~\bibnamefont {{Akutsu}}}, \bibinfo {author} {\bibfnamefont {M.}~\bibnamefont {{Ando}}}, \emph {et~al.},\ }\bibfield  {title} {\bibinfo {title} {{KAGRA: 2.5 generation interferometric gravitational wave detector}},\ }\href {https://doi.org/10.1038/s41550-018-0658-y} {\bibfield  {journal} {\bibinfo  {journal} {Nature Astronomy}\ }\textbf {\bibinfo {volume} {3}},\ \bibinfo {pages} {35} (\bibinfo {year} {2019})},\ \Eprint {https://arxiv.org/abs/1811.08079} {arXiv:1811.08079 [gr-qc]} \BibitemShut {NoStop}%
\bibitem [{\citenamefont {{Xu}}\ \emph {et~al.}(2023)\citenamefont {{Xu}}, \citenamefont {{Chen}}, \citenamefont {{Guo}} \emph {et~al.}}]{2023RAA....23g5024X}%
  \BibitemOpen
  \bibfield  {author} {\bibinfo {author} {\bibfnamefont {H.}~\bibnamefont {{Xu}}}, \bibinfo {author} {\bibfnamefont {S.}~\bibnamefont {{Chen}}}, \bibinfo {author} {\bibfnamefont {Y.}~\bibnamefont {{Guo}}}, \emph {et~al.},\ }\bibfield  {title} {\bibinfo {title} {{Searching for the Nano-Hertz Stochastic Gravitational Wave Background with the Chinese Pulsar Timing Array Data Release I}},\ }\href {https://doi.org/10.1088/1674-4527/acdfa5} {\bibfield  {journal} {\bibinfo  {journal} {Research in Astronomy and Astrophysics}\ }\textbf {\bibinfo {volume} {23}},\ \bibinfo {eid} {075024} (\bibinfo {year} {2023})},\ \Eprint {https://arxiv.org/abs/2306.16216} {arXiv:2306.16216 [astro-ph.HE]} \BibitemShut {NoStop}%
\bibitem [{\citenamefont {{EPTA Collaboration}}\ \emph {et~al.}(2023)\citenamefont {{EPTA Collaboration}}, \citenamefont {{InPTA Collaboration}} \emph {et~al.}}]{2023A&A...678A..50E}%
  \BibitemOpen
  \bibfield  {author} {\bibinfo {author} {\bibnamefont {{EPTA Collaboration}}}, \bibinfo {author} {\bibnamefont {{InPTA Collaboration}}}, \emph {et~al.},\ }\bibfield  {title} {\bibinfo {title} {{The second data release from the European Pulsar Timing Array. III. Search for gravitational wave signals}},\ }\href {https://doi.org/10.1051/0004-6361/202346844} {\bibfield  {journal} {\bibinfo  {journal} {Astronomy \& Astrophysics}\ }\textbf {\bibinfo {volume} {678}},\ \bibinfo {eid} {A50} (\bibinfo {year} {2023})},\ \Eprint {https://arxiv.org/abs/2306.16214} {arXiv:2306.16214 [astro-ph.HE]} \BibitemShut {NoStop}%
\bibitem [{\citenamefont {{Agazie}}\ \emph {et~al.}(2023)\citenamefont {{Agazie}}, \citenamefont {{Alam}}, \citenamefont {{Anumarlapudi}} \emph {et~al.}}]{2023ApJ...951L...9A}%
  \BibitemOpen
  \bibfield  {author} {\bibinfo {author} {\bibfnamefont {G.}~\bibnamefont {{Agazie}}}, \bibinfo {author} {\bibfnamefont {M.~F.}\ \bibnamefont {{Alam}}}, \bibinfo {author} {\bibfnamefont {A.}~\bibnamefont {{Anumarlapudi}}}, \emph {et~al.},\ }\bibfield  {title} {\bibinfo {title} {{The NANOGrav 15 yr Data Set: Observations and Timing of 68 Millisecond Pulsars}},\ }\href {https://doi.org/10.3847/2041-8213/acda9a} {\bibfield  {journal} {\bibinfo  {journal} {The Astrophysical Journal Letters}\ }\textbf {\bibinfo {volume} {951}},\ \bibinfo {eid} {L9} (\bibinfo {year} {2023})},\ \Eprint {https://arxiv.org/abs/2306.16217} {arXiv:2306.16217 [astro-ph.HE]} \BibitemShut {NoStop}%
\bibitem [{\citenamefont {{Reardon}}\ \emph {et~al.}(2023)\citenamefont {{Reardon}}, \citenamefont {{Zic}}, \citenamefont {{Shannon}} \emph {et~al.}}]{2023ApJ...951L...6R}%
  \BibitemOpen
  \bibfield  {author} {\bibinfo {author} {\bibfnamefont {D.~J.}\ \bibnamefont {{Reardon}}}, \bibinfo {author} {\bibfnamefont {A.}~\bibnamefont {{Zic}}}, \bibinfo {author} {\bibfnamefont {R.~M.}\ \bibnamefont {{Shannon}}}, \emph {et~al.},\ }\bibfield  {title} {\bibinfo {title} {{Search for an Isotropic Gravitational-wave Background with the Parkes Pulsar Timing Array}},\ }\href {https://doi.org/10.3847/2041-8213/acdd02} {\bibfield  {journal} {\bibinfo  {journal} {The Astrophysical Journal Letters}\ }\textbf {\bibinfo {volume} {951}},\ \bibinfo {eid} {L6} (\bibinfo {year} {2023})},\ \Eprint {https://arxiv.org/abs/2306.16215} {arXiv:2306.16215 [astro-ph.HE]} \BibitemShut {NoStop}%
\bibitem [{\citenamefont {{Arca Sedda}}\ \emph {et~al.}(2020)\citenamefont {{Arca Sedda}}, \citenamefont {{Berry}}, \citenamefont {{Jani}} \emph {et~al.}}]{2020CQGra..37u5011A}%
  \BibitemOpen
  \bibfield  {author} {\bibinfo {author} {\bibfnamefont {M.}~\bibnamefont {{Arca Sedda}}}, \bibinfo {author} {\bibfnamefont {C.~P.~L.}\ \bibnamefont {{Berry}}}, \bibinfo {author} {\bibfnamefont {K.}~\bibnamefont {{Jani}}}, \emph {et~al.},\ }\bibfield  {title} {\bibinfo {title} {{The missing link in gravitational-wave astronomy: discoveries waiting in the decihertz range}},\ }\href {https://doi.org/10.1088/1361-6382/abb5c1} {\bibfield  {journal} {\bibinfo  {journal} {Classical and Quantum Gravity}\ }\textbf {\bibinfo {volume} {37}},\ \bibinfo {eid} {215011} (\bibinfo {year} {2020})},\ \Eprint {https://arxiv.org/abs/1908.11375} {arXiv:1908.11375 [gr-qc]} \BibitemShut {NoStop}%
\bibitem [{\citenamefont {{Colpi}}\ \emph {et~al.}(2024)\citenamefont {{Colpi}}, \citenamefont {{Danzmann}} \emph {et~al.}}]{2024arXiv240207571C}%
  \BibitemOpen
  \bibfield  {author} {\bibinfo {author} {\bibfnamefont {M.}~\bibnamefont {{Colpi}}}, \bibinfo {author} {\bibfnamefont {K.}~\bibnamefont {{Danzmann}}}, \emph {et~al.},\ }\bibfield  {title} {\bibinfo {title} {{LISA Definition Study Report}},\ }\href {https://doi.org/10.48550/arXiv.2402.07571} {\bibfield  {journal} {\bibinfo  {journal} {arXiv e-prints}\ ,\ \bibinfo {eid} {arXiv:2402.07571}} (\bibinfo {year} {2024})},\ \Eprint {https://arxiv.org/abs/2402.07571} {arXiv:2402.07571 [astro-ph.CO]} \BibitemShut {NoStop}%
\bibitem [{\citenamefont {{Luo}}\ \emph {et~al.}(2016)\citenamefont {{Luo}}, \citenamefont {{Chen}}, \citenamefont {{Duan}} \emph {et~al.}}]{2016CQGra..33c5010L}%
  \BibitemOpen
  \bibfield  {author} {\bibinfo {author} {\bibfnamefont {J.}~\bibnamefont {{Luo}}}, \bibinfo {author} {\bibfnamefont {L.-S.}\ \bibnamefont {{Chen}}}, \bibinfo {author} {\bibfnamefont {H.-Z.}\ \bibnamefont {{Duan}}}, \emph {et~al.},\ }\bibfield  {title} {\bibinfo {title} {{TianQin: a space-borne gravitational wave detector}},\ }\href {https://doi.org/10.1088/0264-9381/33/3/035010} {\bibfield  {journal} {\bibinfo  {journal} {Classical and Quantum Gravity}\ }\textbf {\bibinfo {volume} {33}},\ \bibinfo {eid} {035010} (\bibinfo {year} {2016})},\ \Eprint {https://arxiv.org/abs/1512.02076} {arXiv:1512.02076 [astro-ph.IM]} \BibitemShut {NoStop}%
\bibitem [{\citenamefont {{Luo}}\ \emph {et~al.}(2021)\citenamefont {{Luo}}, \citenamefont {{Wang}}, \citenamefont {{Wu}}, \citenamefont {{Hu}},\ and\ \citenamefont {{Jin}}}]{2021PTEP.2021eA108L}%
  \BibitemOpen
  \bibfield  {author} {\bibinfo {author} {\bibfnamefont {Z.}~\bibnamefont {{Luo}}}, \bibinfo {author} {\bibfnamefont {Y.}~\bibnamefont {{Wang}}}, \bibinfo {author} {\bibfnamefont {Y.}~\bibnamefont {{Wu}}}, \bibinfo {author} {\bibfnamefont {W.}~\bibnamefont {{Hu}}},\ and\ \bibinfo {author} {\bibfnamefont {G.}~\bibnamefont {{Jin}}},\ }\bibfield  {title} {\bibinfo {title} {{The Taiji program: A concise overview}},\ }\href {https://doi.org/10.1093/ptep/ptaa083} {\bibfield  {journal} {\bibinfo  {journal} {Progress of Theoretical and Experimental Physics}\ }\textbf {\bibinfo {volume} {2021}},\ \bibinfo {eid} {05A108} (\bibinfo {year} {2021})}\BibitemShut {NoStop}%
\bibitem [{\citenamefont {{Kawamura}}\ \emph {et~al.}(2011)\citenamefont {{Kawamura}}, \citenamefont {{Ando}}, \citenamefont {{Seto}} \emph {et~al.}}]{2011CQGra..28i4011K}%
  \BibitemOpen
  \bibfield  {author} {\bibinfo {author} {\bibfnamefont {S.}~\bibnamefont {{Kawamura}}}, \bibinfo {author} {\bibfnamefont {M.}~\bibnamefont {{Ando}}}, \bibinfo {author} {\bibfnamefont {N.}~\bibnamefont {{Seto}}}, \emph {et~al.},\ }\bibfield  {title} {\bibinfo {title} {{The Japanese space gravitational wave antenna: DECIGO}},\ }\href {https://doi.org/10.1088/0264-9381/28/9/094011} {\bibfield  {journal} {\bibinfo  {journal} {Classical and Quantum Gravity}\ }\textbf {\bibinfo {volume} {28}},\ \bibinfo {eid} {094011} (\bibinfo {year} {2011})}\BibitemShut {NoStop}%
\bibitem [{\citenamefont {{Ni}}(2022)}]{2022IJMPD..3150039N}%
  \BibitemOpen
  \bibfield  {author} {\bibinfo {author} {\bibfnamefont {W.-T.}\ \bibnamefont {{Ni}}},\ }\bibfield  {title} {\bibinfo {title} {{Core noise and GW sensitivities of AMIGO}},\ }\href {https://doi.org/10.1142/S0218271822500390} {\bibfield  {journal} {\bibinfo  {journal} {International Journal of Modern Physics D}\ }\textbf {\bibinfo {volume} {31}},\ \bibinfo {eid} {2250039} (\bibinfo {year} {2022})},\ \Eprint {https://arxiv.org/abs/2106.12432} {arXiv:2106.12432 [gr-qc]} \BibitemShut {NoStop}%
\bibitem [{\citenamefont {{Weber}}(1960)}]{1960PhRv..117..306W}%
  \BibitemOpen
  \bibfield  {author} {\bibinfo {author} {\bibfnamefont {J.}~\bibnamefont {{Weber}}},\ }\bibfield  {title} {\bibinfo {title} {{Detection and Generation of Gravitational Waves}},\ }\href {https://doi.org/10.1103/PhysRev.117.306} {\bibfield  {journal} {\bibinfo  {journal} {Physical Review}\ }\textbf {\bibinfo {volume} {117}},\ \bibinfo {pages} {306} (\bibinfo {year} {1960})}\BibitemShut {NoStop}%
\bibitem [{\citenamefont {{Weber}}(1968)}]{1968PhT....21d..34W}%
  \BibitemOpen
  \bibfield  {author} {\bibinfo {author} {\bibfnamefont {J.}~\bibnamefont {{Weber}}},\ }\bibfield  {title} {\bibinfo {title} {{Gravitational Waves}},\ }\href {https://doi.org/10.1063/1.3034919} {\bibfield  {journal} {\bibinfo  {journal} {Physics Today}\ }\textbf {\bibinfo {volume} {21}},\ \bibinfo {pages} {34} (\bibinfo {year} {1968})}\BibitemShut {NoStop}%
\bibitem [{\citenamefont {{Dyson}}(1969)}]{1969ApJ...156..529D}%
  \BibitemOpen
  \bibfield  {author} {\bibinfo {author} {\bibfnamefont {F.~J.}\ \bibnamefont {{Dyson}}},\ }\bibfield  {title} {\bibinfo {title} {{Seismic Response of the Earth to a Gravitational Wave in the 1-Hz Band}},\ }\href {https://doi.org/10.1086/149986} {\bibfield  {journal} {\bibinfo  {journal} {\apj}\ }\textbf {\bibinfo {volume} {156}},\ \bibinfo {pages} {529} (\bibinfo {year} {1969})}\BibitemShut {NoStop}%
\bibitem [{\citenamefont {{Lognonn{\'e}}}\ \emph {et~al.}(2009)\citenamefont {{Lognonn{\'e}}}, \citenamefont {{Le Feuvre}}, \citenamefont {{Johnson}},\ and\ \citenamefont {{Weber}}}]{2009JGRE..11412003L}%
  \BibitemOpen
  \bibfield  {author} {\bibinfo {author} {\bibfnamefont {P.}~\bibnamefont {{Lognonn{\'e}}}}, \bibinfo {author} {\bibfnamefont {M.}~\bibnamefont {{Le Feuvre}}}, \bibinfo {author} {\bibfnamefont {C.~L.}\ \bibnamefont {{Johnson}}},\ and\ \bibinfo {author} {\bibfnamefont {R.~C.}\ \bibnamefont {{Weber}}},\ }\bibfield  {title} {\bibinfo {title} {{Moon meteoritic seismic hum: Steady state prediction}},\ }\href {https://doi.org/10.1029/2008JE003294} {\bibfield  {journal} {\bibinfo  {journal} {Journal of Geophysical Research (Planets)}\ }\textbf {\bibinfo {volume} {114}},\ \bibinfo {eid} {E12003} (\bibinfo {year} {2009})}\BibitemShut {NoStop}%
\bibitem [{\citenamefont {{Ben-Menahem}}(1983)}]{1983NCimC...6...49B}%
  \BibitemOpen
  \bibfield  {author} {\bibinfo {author} {\bibfnamefont {A.}~\bibnamefont {{Ben-Menahem}}},\ }\bibfield  {title} {\bibinfo {title} {{Excitation of the earth's eigenvibrations by gravitational radiation from astrophysical sources.}},\ }\href {https://doi.org/10.1007/BF02511372} {\bibfield  {journal} {\bibinfo  {journal} {Nuovo Cimento C Geophysics Space Physics C}\ }\textbf {\bibinfo {volume} {6}},\ \bibinfo {pages} {49} (\bibinfo {year} {1983})}\BibitemShut {NoStop}%
\bibitem [{\citenamefont {{Bianchi}}\ \emph {et~al.}(1996)\citenamefont {{Bianchi}}, \citenamefont {{Coccia}}, \citenamefont {{Colacino}}, \citenamefont {{Fafone}},\ and\ \citenamefont {{Fucito}}}]{1996CQGra..13.2865B}%
  \BibitemOpen
  \bibfield  {author} {\bibinfo {author} {\bibfnamefont {M.}~\bibnamefont {{Bianchi}}}, \bibinfo {author} {\bibfnamefont {E.}~\bibnamefont {{Coccia}}}, \bibinfo {author} {\bibfnamefont {C.~N.}\ \bibnamefont {{Colacino}}}, \bibinfo {author} {\bibfnamefont {V.}~\bibnamefont {{Fafone}}},\ and\ \bibinfo {author} {\bibfnamefont {F.}~\bibnamefont {{Fucito}}},\ }\bibfield  {title} {\bibinfo {title} {{Testing theories of gravity with a spherical gravitational wave detector}},\ }\href {https://doi.org/10.1088/0264-9381/13/11/003} {\bibfield  {journal} {\bibinfo  {journal} {Classical and Quantum Gravity}\ }\textbf {\bibinfo {volume} {13}},\ \bibinfo {pages} {2865} (\bibinfo {year} {1996})},\ \Eprint {https://arxiv.org/abs/gr-qc/9604026} {arXiv:gr-qc/9604026 [gr-qc]} \BibitemShut {NoStop}%
\bibitem [{\citenamefont {{Majstorovi{\'c}}}\ \emph {et~al.}(2019)\citenamefont {{Majstorovi{\'c}}}, \citenamefont {{Rosat}},\ and\ \citenamefont {{Rogister}}}]{2019PhRvD.100d4048M}%
  \BibitemOpen
  \bibfield  {author} {\bibinfo {author} {\bibfnamefont {J.}~\bibnamefont {{Majstorovi{\'c}}}}, \bibinfo {author} {\bibfnamefont {S.}~\bibnamefont {{Rosat}}},\ and\ \bibinfo {author} {\bibfnamefont {Y.}~\bibnamefont {{Rogister}}},\ }\bibfield  {title} {\bibinfo {title} {{Earth's spheroidal motion induced by a gravitational wave in flat spacetime}},\ }\href {https://doi.org/10.1103/PhysRevD.100.044048} {\bibfield  {journal} {\bibinfo  {journal} {\prd}\ }\textbf {\bibinfo {volume} {100}},\ \bibinfo {eid} {044048} (\bibinfo {year} {2019})}\BibitemShut {NoStop}%
\bibitem [{\citenamefont {Yan}\ \emph {et~al.}(2024)\citenamefont {Yan}, \citenamefont {Chen}, \citenamefont {Zhang}, \citenamefont {Zhang}, \citenamefont {Wang},\ and\ \citenamefont {Shao}}]{PhysRevD.109.064092}%
  \BibitemOpen
  \bibfield  {author} {\bibinfo {author} {\bibfnamefont {H.}~\bibnamefont {Yan}}, \bibinfo {author} {\bibfnamefont {X.}~\bibnamefont {Chen}}, \bibinfo {author} {\bibfnamefont {J.}~\bibnamefont {Zhang}}, \bibinfo {author} {\bibfnamefont {F.}~\bibnamefont {Zhang}}, \bibinfo {author} {\bibfnamefont {M.}~\bibnamefont {Wang}},\ and\ \bibinfo {author} {\bibfnamefont {L.}~\bibnamefont {Shao}},\ }\bibfield  {title} {\bibinfo {title} {Toward a consistent calculation of the lunar response to gravitational waves},\ }\href {https://doi.org/10.1103/PhysRevD.109.064092} {\bibfield  {journal} {\bibinfo  {journal} {Phys. Rev. D}\ }\textbf {\bibinfo {volume} {109}},\ \bibinfo {pages} {064092} (\bibinfo {year} {2024})}\BibitemShut {NoStop}%
\bibitem [{\citenamefont {{Majstorovi{\'c}}}\ \emph {et~al.}(2025)\citenamefont {{Majstorovi{\'c}}}, \citenamefont {{Vidal}},\ and\ \citenamefont {{Lognonn{\'e}}}}]{2025PhRvD.111d4061M}%
  \BibitemOpen
  \bibfield  {author} {\bibinfo {author} {\bibfnamefont {J.}~\bibnamefont {{Majstorovi{\'c}}}}, \bibinfo {author} {\bibfnamefont {L.}~\bibnamefont {{Vidal}}},\ and\ \bibinfo {author} {\bibfnamefont {P.}~\bibnamefont {{Lognonn{\'e}}}},\ }\bibfield  {title} {\bibinfo {title} {{Modeling lunar response to gravitational waves using normal-mode approach and tidal forcing}},\ }\href {https://doi.org/10.1103/PhysRevD.111.044061} {\bibfield  {journal} {\bibinfo  {journal} {\prd}\ }\textbf {\bibinfo {volume} {111}},\ \bibinfo {eid} {044061} (\bibinfo {year} {2025})},\ \Eprint {https://arxiv.org/abs/2411.09559} {arXiv:2411.09559 [gr-qc]} \BibitemShut {NoStop}%
\bibitem [{\citenamefont {{Bi}}\ and\ \citenamefont {{Harms}}(2024)}]{2024PhRvD.110f4025B}%
  \BibitemOpen
  \bibfield  {author} {\bibinfo {author} {\bibfnamefont {X.}~\bibnamefont {{Bi}}}\ and\ \bibinfo {author} {\bibfnamefont {J.}~\bibnamefont {{Harms}}},\ }\bibfield  {title} {\bibinfo {title} {{Response of the Moon to gravitational waves}},\ }\href {https://doi.org/10.1103/PhysRevD.110.064025} {\bibfield  {journal} {\bibinfo  {journal} {\prd}\ }\textbf {\bibinfo {volume} {110}},\ \bibinfo {eid} {064025} (\bibinfo {year} {2024})},\ \Eprint {https://arxiv.org/abs/2403.05118} {arXiv:2403.05118 [gr-qc]} \BibitemShut {NoStop}%
\bibitem [{\citenamefont {{Smith}}\ \emph {et~al.}(2010)\citenamefont {{Smith}}, \citenamefont {{Zuber}} \emph {et~al.}}]{2010GeoRL..3718204S}%
  \BibitemOpen
  \bibfield  {author} {\bibinfo {author} {\bibfnamefont {D.~E.}\ \bibnamefont {{Smith}}}, \bibinfo {author} {\bibfnamefont {M.~T.}\ \bibnamefont {{Zuber}}}, \emph {et~al.},\ }\bibfield  {title} {\bibinfo {title} {{Initial observations from the Lunar Orbiter Laser Altimeter (LOLA)}},\ }\href {https://doi.org/10.1029/2010GL043751} {\bibfield  {journal} {\bibinfo  {journal} {Geophysical Research Letters}\ }\textbf {\bibinfo {volume} {37}},\ \bibinfo {eid} {L18204} (\bibinfo {year} {2010})}\BibitemShut {NoStop}%
\bibitem [{\citenamefont {{Wieczorek}}\ \emph {et~al.}(2013)\citenamefont {{Wieczorek}}, \citenamefont {{Neumann}}, \citenamefont {{Nimmo}} \emph {et~al.}}]{2013Sci...339..671W}%
  \BibitemOpen
  \bibfield  {author} {\bibinfo {author} {\bibfnamefont {M.~A.}\ \bibnamefont {{Wieczorek}}}, \bibinfo {author} {\bibfnamefont {G.~A.}\ \bibnamefont {{Neumann}}}, \bibinfo {author} {\bibfnamefont {F.}~\bibnamefont {{Nimmo}}}, \emph {et~al.},\ }\bibfield  {title} {\bibinfo {title} {{The Crust of the Moon as Seen by GRAIL}},\ }\href {https://doi.org/10.1126/science.1231530} {\bibfield  {journal} {\bibinfo  {journal} {Science}\ }\textbf {\bibinfo {volume} {339}},\ \bibinfo {pages} {671} (\bibinfo {year} {2013})}\BibitemShut {NoStop}%
\bibitem [{\citenamefont {{Huang}}\ \emph {et~al.}(2022)\citenamefont {{Huang}}, \citenamefont {{Soderblom}}, \citenamefont {{Minton}}, \citenamefont {{Hirabayashi}},\ and\ \citenamefont {{Melosh}}}]{huang2022bombardment}%
  \BibitemOpen
  \bibfield  {author} {\bibinfo {author} {\bibfnamefont {Y.~H.}\ \bibnamefont {{Huang}}}, \bibinfo {author} {\bibfnamefont {J.~M.}\ \bibnamefont {{Soderblom}}}, \bibinfo {author} {\bibfnamefont {D.~A.}\ \bibnamefont {{Minton}}}, \bibinfo {author} {\bibfnamefont {M.}~\bibnamefont {{Hirabayashi}}},\ and\ \bibinfo {author} {\bibfnamefont {H.~J.}\ \bibnamefont {{Melosh}}},\ }\bibfield  {title} {\bibinfo {title} {Bombardment history of the moon constrained by crustal porosity},\ }\href@noop {} {\bibfield  {journal} {\bibinfo  {journal} {Nature Geoscience}\ }\textbf {\bibinfo {volume} {15}},\ \bibinfo {pages} {531} (\bibinfo {year} {2022})}\BibitemShut {NoStop}%
\bibitem [{\citenamefont {{Wieczorek}}\ and\ \citenamefont {{Phillips}}(1998)}]{wieczorek1998potential}%
  \BibitemOpen
  \bibfield  {author} {\bibinfo {author} {\bibfnamefont {M.~A.}\ \bibnamefont {{Wieczorek}}}\ and\ \bibinfo {author} {\bibfnamefont {R.~J.}\ \bibnamefont {{Phillips}}},\ }\bibfield  {title} {\bibinfo {title} {Potential anomalies on a sphere: Applications to the thickness of the lunar crust},\ }\href@noop {} {\bibfield  {journal} {\bibinfo  {journal} {Journal of Geophysical Research: Planets}\ }\textbf {\bibinfo {volume} {103}},\ \bibinfo {pages} {1715} (\bibinfo {year} {1998})}\BibitemShut {NoStop}%
\bibitem [{\citenamefont {{Woodhouse}}(1980)}]{1980GeoJ...61..261W}%
  \BibitemOpen
  \bibfield  {author} {\bibinfo {author} {\bibfnamefont {J.~H.}\ \bibnamefont {{Woodhouse}}},\ }\bibfield  {title} {\bibinfo {title} {{The coupling and attenuation of nearly resonant multiplets in the Earth's free oscillation spectrum}},\ }\href {https://doi.org/10.1111/j.1365-246X.1980.tb04317.x} {\bibfield  {journal} {\bibinfo  {journal} {Geophysical Journal}\ }\textbf {\bibinfo {volume} {61}},\ \bibinfo {pages} {261} (\bibinfo {year} {1980})}\BibitemShut {NoStop}%
\bibitem [{\citenamefont {{Dahlen}}\ and\ \citenamefont {{Tromp}}(1999)}]{1999tgs..book.....D}%
  \BibitemOpen
  \bibfield  {author} {\bibinfo {author} {\bibfnamefont {F.~A.}\ \bibnamefont {{Dahlen}}}\ and\ \bibinfo {author} {\bibfnamefont {J.}~\bibnamefont {{Tromp}}},\ }\href@noop {} {\emph {\bibinfo {title} {{Theoretical Global Seismology}}}}\ (\bibinfo  {publisher} {Princeton University Press},\ \bibinfo {year} {1999})\BibitemShut {NoStop}%
\bibitem [{\citenamefont {{Woodhouse}}\ and\ \citenamefont {{Dahlen}}(1978)}]{1978GeoJ...53..335W}%
  \BibitemOpen
  \bibfield  {author} {\bibinfo {author} {\bibfnamefont {J.~H.}\ \bibnamefont {{Woodhouse}}}\ and\ \bibinfo {author} {\bibfnamefont {F.~A.}\ \bibnamefont {{Dahlen}}},\ }\bibfield  {title} {\bibinfo {title} {{The effect of a general aspherical perturbation on the free oscillations of the earth.}},\ }\href {https://doi.org/10.1111/j.1365-246X.1978.tb03746.x} {\bibfield  {journal} {\bibinfo  {journal} {Geophysical Journal International}\ }\textbf {\bibinfo {volume} {53}},\ \bibinfo {pages} {335} (\bibinfo {year} {1978})}\BibitemShut {NoStop}%
\bibitem [{\citenamefont {{Masters}}\ \emph {et~al.}(1983)\citenamefont {{Masters}}, \citenamefont {{Park}},\ and\ \citenamefont {{Gilbert}}}]{1983JGR....8810285M}%
  \BibitemOpen
  \bibfield  {author} {\bibinfo {author} {\bibfnamefont {G.}~\bibnamefont {{Masters}}}, \bibinfo {author} {\bibfnamefont {J.}~\bibnamefont {{Park}}},\ and\ \bibinfo {author} {\bibfnamefont {F.}~\bibnamefont {{Gilbert}}},\ }\bibfield  {title} {\bibinfo {title} {{Observations of coupled spheroidal and toroidal modes}},\ }\href {https://doi.org/10.1029/JB088iB12p10285} {\bibfield  {journal} {\bibinfo  {journal} {Journal of Geophysical Research}\ }\textbf {\bibinfo {volume} {88}},\ \bibinfo {pages} {10285} (\bibinfo {year} {1983})}\BibitemShut {NoStop}%
\bibitem [{\citenamefont {{Deuss}}\ and\ \citenamefont {{Woodhouse}}(2001)}]{2001GeoJI.146..833D}%
  \BibitemOpen
  \bibfield  {author} {\bibinfo {author} {\bibfnamefont {A.}~\bibnamefont {{Deuss}}}\ and\ \bibinfo {author} {\bibfnamefont {J.~H.}\ \bibnamefont {{Woodhouse}}},\ }\bibfield  {title} {\bibinfo {title} {{Theoretical free-oscillation spectra: the importance of wide band coupling}},\ }\href {https://doi.org/10.1046/j.0956-540X.2001.00502.x} {\bibfield  {journal} {\bibinfo  {journal} {Geophysical Journal International}\ }\textbf {\bibinfo {volume} {146}},\ \bibinfo {pages} {833} (\bibinfo {year} {2001})}\BibitemShut {NoStop}%
\bibitem [{\citenamefont {Chen}(2026)}]{Chen2026-be}%
  \BibitemOpen
  \bibfield  {author} {\bibinfo {author} {\bibfnamefont {X.}~\bibnamefont {Chen}},\ }\bibfield  {title} {\bibinfo {title} {The moon as a gateway to discovery: how lunar gravitational-wave detection advances science across disciplines},\ }\href@noop {} {\bibfield  {journal} {\bibinfo  {journal} {npj Space Exploration}\ }\textbf {\bibinfo {volume} {2}} (\bibinfo {year} {2026})}\BibitemShut {NoStop}%
\bibitem [{\citenamefont {{Harms}}\ \emph {et~al.}(2021)\citenamefont {{Harms}}, \citenamefont {{Ambrosino}}, \citenamefont {{Angelini}} \emph {et~al.}}]{2021ApJ...910....1H}%
  \BibitemOpen
  \bibfield  {author} {\bibinfo {author} {\bibfnamefont {J.}~\bibnamefont {{Harms}}}, \bibinfo {author} {\bibfnamefont {F.}~\bibnamefont {{Ambrosino}}}, \bibinfo {author} {\bibfnamefont {L.}~\bibnamefont {{Angelini}}}, \emph {et~al.},\ }\bibfield  {title} {\bibinfo {title} {{Lunar Gravitational-wave Antenna}},\ }\href {https://doi.org/10.3847/1538-4357/abe5a7} {\bibfield  {journal} {\bibinfo  {journal} {\apj}\ }\textbf {\bibinfo {volume} {910}},\ \bibinfo {eid} {1} (\bibinfo {year} {2021})},\ \Eprint {https://arxiv.org/abs/2010.13726} {arXiv:2010.13726 [gr-qc]} \BibitemShut {NoStop}%
\bibitem [{\citenamefont {{Branchesi}}\ \emph {et~al.}(2023)\citenamefont {{Branchesi}}, \citenamefont {{Falanga}}, \citenamefont {{Harms}} \emph {et~al.}}]{2023SSRv..219...67B}%
  \BibitemOpen
  \bibfield  {author} {\bibinfo {author} {\bibfnamefont {M.}~\bibnamefont {{Branchesi}}}, \bibinfo {author} {\bibfnamefont {M.}~\bibnamefont {{Falanga}}}, \bibinfo {author} {\bibfnamefont {J.}~\bibnamefont {{Harms}}}, \emph {et~al.},\ }\bibfield  {title} {\bibinfo {title} {{Lunar Gravitational-Wave Detection}},\ }\href {https://doi.org/10.1007/s11214-023-01015-4} {\bibfield  {journal} {\bibinfo  {journal} {Space Science Reviews}\ }\textbf {\bibinfo {volume} {219}},\ \bibinfo {eid} {67} (\bibinfo {year} {2023})}\BibitemShut {NoStop}%
\bibitem [{\citenamefont {{Jani}}\ \emph {et~al.}(2025)\citenamefont {{Jani}}, \citenamefont {{Abernathy}}, \citenamefont {{Berti}} \emph {et~al.}}]{2025arXiv250811631J}%
  \BibitemOpen
  \bibfield  {author} {\bibinfo {author} {\bibfnamefont {K.}~\bibnamefont {{Jani}}}, \bibinfo {author} {\bibfnamefont {M.}~\bibnamefont {{Abernathy}}}, \bibinfo {author} {\bibfnamefont {E.}~\bibnamefont {{Berti}}}, \emph {et~al.},\ }\bibfield  {title} {\bibinfo {title} {{Laser Interferometer Lunar Antenna (LILA): Advancing the U.S. Priorities in Gravitational-wave and Lunar Science}},\ }\href {https://doi.org/10.48550/arXiv.2508.11631} {\bibfield  {journal} {\bibinfo  {journal} {arXiv e-prints}\ ,\ \bibinfo {eid} {arXiv:2508.11631}} (\bibinfo {year} {2025})},\ \Eprint {https://arxiv.org/abs/2508.11631} {arXiv:2508.11631 [gr-qc]} \BibitemShut {NoStop}%
\bibitem [{\citenamefont {{Paik}}\ and\ \citenamefont {{Venkateswara}}(2009)}]{2009AdSpR..43..167P}%
  \BibitemOpen
  \bibfield  {author} {\bibinfo {author} {\bibfnamefont {H.~J.}\ \bibnamefont {{Paik}}}\ and\ \bibinfo {author} {\bibfnamefont {K.~Y.}\ \bibnamefont {{Venkateswara}}},\ }\bibfield  {title} {\bibinfo {title} {{Gravitational wave detection on the Moon and the moons of Mars}},\ }\href {https://doi.org/10.1016/j.asr.2008.04.010} {\bibfield  {journal} {\bibinfo  {journal} {Advances in Space Research}\ }\textbf {\bibinfo {volume} {43}},\ \bibinfo {pages} {167} (\bibinfo {year} {2009})}\BibitemShut {NoStop}%
\bibitem [{\citenamefont {{Amaro-Seoane}}\ \emph {et~al.}(2021)\citenamefont {{Amaro-Seoane}}, \citenamefont {{Bischof}}, \citenamefont {{Carter}}, \citenamefont {{Hartig}},\ and\ \citenamefont {{Wilken}}}]{2021CQGra..38l5008A}%
  \BibitemOpen
  \bibfield  {author} {\bibinfo {author} {\bibfnamefont {P.}~\bibnamefont {{Amaro-Seoane}}}, \bibinfo {author} {\bibfnamefont {L.}~\bibnamefont {{Bischof}}}, \bibinfo {author} {\bibfnamefont {J.~J.}\ \bibnamefont {{Carter}}}, \bibinfo {author} {\bibfnamefont {M.-S.}\ \bibnamefont {{Hartig}}},\ and\ \bibinfo {author} {\bibfnamefont {D.}~\bibnamefont {{Wilken}}},\ }\bibfield  {title} {\bibinfo {title} {{LION: laser interferometer on the moon}},\ }\href {https://doi.org/10.1088/1361-6382/abf441} {\bibfield  {journal} {\bibinfo  {journal} {Classical and Quantum Gravity}\ }\textbf {\bibinfo {volume} {38}},\ \bibinfo {eid} {125008} (\bibinfo {year} {2021})},\ \Eprint {https://arxiv.org/abs/2012.10443} {arXiv:2012.10443 [astro-ph.IM]} \BibitemShut {NoStop}%
\bibitem [{\citenamefont {{Li}}\ \emph {et~al.}(2023)\citenamefont {{Li}}, \citenamefont {{Liu}}, \citenamefont {{Pan}}, \citenamefont {{Wang}}, \citenamefont {{Cao}}, \citenamefont {{Wang}}, \citenamefont {{Zhang}}, \citenamefont {{Zhang}},\ and\ \citenamefont {{Zhu}}}]{2023SCPMA..6609513L}%
  \BibitemOpen
  \bibfield  {author} {\bibinfo {author} {\bibfnamefont {J.}~\bibnamefont {{Li}}}, \bibinfo {author} {\bibfnamefont {F.}~\bibnamefont {{Liu}}}, \bibinfo {author} {\bibfnamefont {Y.}~\bibnamefont {{Pan}}}, \bibinfo {author} {\bibfnamefont {Z.}~\bibnamefont {{Wang}}}, \bibinfo {author} {\bibfnamefont {M.}~\bibnamefont {{Cao}}}, \bibinfo {author} {\bibfnamefont {M.}~\bibnamefont {{Wang}}}, \bibinfo {author} {\bibfnamefont {F.}~\bibnamefont {{Zhang}}}, \bibinfo {author} {\bibfnamefont {J.}~\bibnamefont {{Zhang}}},\ and\ \bibinfo {author} {\bibfnamefont {Z.-H.}\ \bibnamefont {{Zhu}}},\ }\bibfield  {title} {\bibinfo {title} {{Detecting gravitational wave with an interferometric seismometer array on lunar nearside}},\ }\href {https://doi.org/10.1007/s11433-023-2179-9} {\bibfield  {journal} {\bibinfo  {journal} {Science China Physics, Mechanics, and Astronomy}\ }\textbf {\bibinfo {volume} {66}},\ \bibinfo {eid} {109513} (\bibinfo {year} {2023})}\BibitemShut {NoStop}%
\bibitem [{\citenamefont {{Zhang}}\ \emph {et~al.}(2025)\citenamefont {{Zhang}}, \citenamefont {{Yan}}, \citenamefont {{Chen}},\ and\ \citenamefont {{Zhang}}}]{2025PhRvD.111f3014Z}%
  \BibitemOpen
  \bibfield  {author} {\bibinfo {author} {\bibfnamefont {L.}~\bibnamefont {{Zhang}}}, \bibinfo {author} {\bibfnamefont {H.}~\bibnamefont {{Yan}}}, \bibinfo {author} {\bibfnamefont {X.}~\bibnamefont {{Chen}}},\ and\ \bibinfo {author} {\bibfnamefont {J.}~\bibnamefont {{Zhang}}},\ }\bibfield  {title} {\bibinfo {title} {{2D numerical simulation of lunar response to gravitational waves using finite element method}},\ }\href {https://doi.org/10.1103/PhysRevD.111.063014} {\bibfield  {journal} {\bibinfo  {journal} {\prd}\ }\textbf {\bibinfo {volume} {111}},\ \bibinfo {eid} {063014} (\bibinfo {year} {2025})},\ \Eprint {https://arxiv.org/abs/2412.17898} {arXiv:2412.17898 [astro-ph.EP]} \BibitemShut {NoStop}%
\bibitem [{\citenamefont {{Zhang}}\ \emph {et~al.}(2026)\citenamefont {{Zhang}}, \citenamefont {{Yan}}, \citenamefont {{Zhang}},\ and\ \citenamefont {{Chen}}}]{2026PhRvD.113b3031Z}%
  \BibitemOpen
  \bibfield  {author} {\bibinfo {author} {\bibfnamefont {L.}~\bibnamefont {{Zhang}}}, \bibinfo {author} {\bibfnamefont {H.}~\bibnamefont {{Yan}}}, \bibinfo {author} {\bibfnamefont {J.}~\bibnamefont {{Zhang}}},\ and\ \bibinfo {author} {\bibfnamefont {X.}~\bibnamefont {{Chen}}},\ }\bibfield  {title} {\bibinfo {title} {{Numerical simulation of lunar response to gravitational waves and its 3D topographic effect using the spectral-element method}},\ }\href {https://doi.org/10.1103/4rrr-w4tm} {\bibfield  {journal} {\bibinfo  {journal} {\prd}\ }\textbf {\bibinfo {volume} {113}},\ \bibinfo {eid} {023031} (\bibinfo {year} {2026})},\ \Eprint {https://arxiv.org/abs/2512.21667} {arXiv:2512.21667 [astro-ph.EP]} \BibitemShut {NoStop}%
\bibitem [{\citenamefont {Komatitsch}\ \emph {et~al.}(2005)\citenamefont {Komatitsch}, \citenamefont {Tsuboi}, \citenamefont {Tromp}, \citenamefont {Levander},\ and\ \citenamefont {Nolet}}]{komatitsch2005spectral}%
  \BibitemOpen
  \bibfield  {author} {\bibinfo {author} {\bibfnamefont {D.}~\bibnamefont {Komatitsch}}, \bibinfo {author} {\bibfnamefont {S.}~\bibnamefont {Tsuboi}}, \bibinfo {author} {\bibfnamefont {J.}~\bibnamefont {Tromp}}, \bibinfo {author} {\bibfnamefont {A.}~\bibnamefont {Levander}},\ and\ \bibinfo {author} {\bibfnamefont {G.}~\bibnamefont {Nolet}},\ }\bibfield  {title} {\bibinfo {title} {{The spectral-element method in seismology}},\ }\href@noop {} {\bibfield  {journal} {\bibinfo  {journal} {Geophysical Monograph-American Geophysical Union}\ }\textbf {\bibinfo {volume} {157}},\ \bibinfo {pages} {205} (\bibinfo {year} {2005})}\BibitemShut {NoStop}%
\bibitem [{\citenamefont {{Zhang}}\ \emph {et~al.}(2020)\citenamefont {{Zhang}}, \citenamefont {{Wang}}, \citenamefont {{Xu}}, \citenamefont {{He}},\ and\ \citenamefont {{Zhang}}}]{zhang2020procedure}%
  \BibitemOpen
  \bibfield  {author} {\bibinfo {author} {\bibfnamefont {L.}~\bibnamefont {{Zhang}}}, \bibinfo {author} {\bibfnamefont {J.}~\bibnamefont {{Wang}}}, \bibinfo {author} {\bibfnamefont {Y.}~\bibnamefont {{Xu}}}, \bibinfo {author} {\bibfnamefont {C.}~\bibnamefont {{He}}},\ and\ \bibinfo {author} {\bibfnamefont {C.}~\bibnamefont {{Zhang}}},\ }\bibfield  {title} {\bibinfo {title} {{A procedure for 3D seismic simulation from rupture to structures by coupling SEM and FEM}},\ }\href@noop {} {\bibfield  {journal} {\bibinfo  {journal} {Bulletin of the Seismological Society of America}\ }\textbf {\bibinfo {volume} {110}},\ \bibinfo {pages} {1134} (\bibinfo {year} {2020})}\BibitemShut {NoStop}%
\bibitem [{sup()}]{supplemental}%
  \BibitemOpen
  \href@noop {} {}\bibinfo {note} {See Supplemental Material at \url{https://...} for the details of the numerical calculation and theoretical formalism, which includes Refs. [56-57].}\BibitemShut {Stop}%
\bibitem [{web(2025)}]{webs}%
  \BibitemOpen
  \href {{https://doi.org/10.5281/zenodo.20309656}} {\bibinfo {title} {See \textit{CrustalAmplify} folder in {https://doi.org/10.5281/zenodo.20309656}}} (\bibinfo {year} {2026})\BibitemShut {NoStop}%
\bibitem [{\citenamefont {Masters}\ \emph {et~al.}(2011)\citenamefont {Masters}, \citenamefont {Woodhouse},\ and\ \citenamefont {Freeman}}]{mineos2011}%
  \BibitemOpen
  \bibfield  {author} {\bibinfo {author} {\bibfnamefont {G.}~\bibnamefont {Masters}}, \bibinfo {author} {\bibfnamefont {J.~H.}\ \bibnamefont {Woodhouse}},\ and\ \bibinfo {author} {\bibfnamefont {G.}~\bibnamefont {Freeman}},\ }\href@noop {} {\bibinfo {title} {Mineos v1.0.2 [software]}},\ \bibinfo {howpublished} {Computational Infrastructure for Geodynamics} (\bibinfo {year} {2011}),\ \bibinfo {note} {\url{https://geodynamics.org/cig}}\BibitemShut {NoStop}%
\bibitem [{\citenamefont {{Yan}}\ \emph {et~al.}(2024)\citenamefont {{Yan}}, \citenamefont {{Chen}}, \citenamefont {{Zhang}}, \citenamefont {{Zhang}}, \citenamefont {{Shao}},\ and\ \citenamefont {{Wang}}}]{2024PhRvD.110d3009Y}%
  \BibitemOpen
  \bibfield  {author} {\bibinfo {author} {\bibfnamefont {H.}~\bibnamefont {{Yan}}}, \bibinfo {author} {\bibfnamefont {X.}~\bibnamefont {{Chen}}}, \bibinfo {author} {\bibfnamefont {J.}~\bibnamefont {{Zhang}}}, \bibinfo {author} {\bibfnamefont {F.}~\bibnamefont {{Zhang}}}, \bibinfo {author} {\bibfnamefont {L.}~\bibnamefont {{Shao}}},\ and\ \bibinfo {author} {\bibfnamefont {M.}~\bibnamefont {{Wang}}},\ }\bibfield  {title} {\bibinfo {title} {{Constraining the stochastic gravitational wave background using the future lunar seismometers}},\ }\href {https://doi.org/10.1103/PhysRevD.110.043009} {\bibfield  {journal} {\bibinfo  {journal} {\prd}\ }\textbf {\bibinfo {volume} {110}},\ \bibinfo {eid} {043009} (\bibinfo {year} {2024})},\ \Eprint {https://arxiv.org/abs/2405.12640} {arXiv:2405.12640 [gr-qc]} \BibitemShut {NoStop}%
\bibitem [{\citenamefont {Zhang}\ \emph {et~al.}(2022{\natexlab{a}})\citenamefont {Zhang}, \citenamefont {Zhang},\ and\ \citenamefont {Mitchell}}]{zhang2022dichotomy}%
  \BibitemOpen
  \bibfield  {author} {\bibinfo {author} {\bibfnamefont {L.}~\bibnamefont {Zhang}}, \bibinfo {author} {\bibfnamefont {J.}~\bibnamefont {Zhang}},\ and\ \bibinfo {author} {\bibfnamefont {R.~N.}\ \bibnamefont {Mitchell}},\ }\bibfield  {title} {\bibinfo {title} {Dichotomy in crustal melting on early mars inferred from antipodal effect},\ }\href@noop {} {\bibfield  {journal} {\bibinfo  {journal} {The Innovation}\ }\textbf {\bibinfo {volume} {3}} (\bibinfo {year} {2022}{\natexlab{a}})}\BibitemShut {NoStop}%
\bibitem [{\citenamefont {Zhang}\ \emph {et~al.}(2022{\natexlab{b}})\citenamefont {Zhang}, \citenamefont {Zhang}, \citenamefont {Zhang}, \citenamefont {Mitchell} \emph {et~al.}}]{ZhangX2022}%
  \BibitemOpen
  \bibfield  {author} {\bibinfo {author} {\bibfnamefont {X.}~\bibnamefont {Zhang}}, \bibinfo {author} {\bibfnamefont {L.}~\bibnamefont {Zhang}}, \bibinfo {author} {\bibfnamefont {J.}~\bibnamefont {Zhang}}, \bibinfo {author} {\bibfnamefont {R.~N.}\ \bibnamefont {Mitchell}}, \emph {et~al.},\ }\bibfield  {title} {\bibinfo {title} {Strong heterogeneity in shallow lunar subsurface detected by {Apollo} seismic data},\ }\href {https://doi.org/10.1029/2022JE007222} {\bibfield  {journal} {\bibinfo  {journal} {Journal of Geophysical Research: Planets}\ }\textbf {\bibinfo {volume} {127}},\ \bibinfo {pages} {e2022JE007222} (\bibinfo {year} {2022}{\natexlab{b}})}\BibitemShut {NoStop}%
\bibitem [{\citenamefont {Onodera}\ \emph {et~al.}(2022)\citenamefont {Onodera}, \citenamefont {Kawamura}, \citenamefont {Tanaka}, \citenamefont {Ishihara},\ and\ \citenamefont {Maeda}}]{Onodera2022}%
  \BibitemOpen
  \bibfield  {author} {\bibinfo {author} {\bibfnamefont {K.}~\bibnamefont {Onodera}}, \bibinfo {author} {\bibfnamefont {T.}~\bibnamefont {Kawamura}}, \bibinfo {author} {\bibfnamefont {S.}~\bibnamefont {Tanaka}}, \bibinfo {author} {\bibfnamefont {Y.}~\bibnamefont {Ishihara}},\ and\ \bibinfo {author} {\bibfnamefont {T.}~\bibnamefont {Maeda}},\ }\bibfield  {title} {\bibinfo {title} {Quantitative evaluation of the lunar seismic scattering and comparison between the {Earth}, {Mars}, and the {Moon}},\ }\href {https://doi.org/10.1029/2022JE007558} {\bibfield  {journal} {\bibinfo  {journal} {Journal of Geophysical Research: Planets}\ }\textbf {\bibinfo {volume} {127}},\ \bibinfo {pages} {e2022JE007558} (\bibinfo {year} {2022})}\BibitemShut {NoStop}%
\bibitem [{\citenamefont {Wiggins}\ \emph {et~al.}(2019)\citenamefont {Wiggins}, \citenamefont {Johnson}, \citenamefont {Bowling}, \citenamefont {Melosh},\ and\ \citenamefont {Silber}}]{Wiggins2019}%
  \BibitemOpen
  \bibfield  {author} {\bibinfo {author} {\bibfnamefont {S.~E.}\ \bibnamefont {Wiggins}}, \bibinfo {author} {\bibfnamefont {B.~C.}\ \bibnamefont {Johnson}}, \bibinfo {author} {\bibfnamefont {T.~J.}\ \bibnamefont {Bowling}}, \bibinfo {author} {\bibfnamefont {H.~J.}\ \bibnamefont {Melosh}},\ and\ \bibinfo {author} {\bibfnamefont {E.~A.}\ \bibnamefont {Silber}},\ }\bibfield  {title} {\bibinfo {title} {Impact fragmentation and the development of the deep lunar megaregolith},\ }\href {https://doi.org/10.1029/2018JE005757} {\bibfield  {journal} {\bibinfo  {journal} {Journal of Geophysical Research: Planets}\ }\textbf {\bibinfo {volume} {124}},\ \bibinfo {pages} {941} (\bibinfo {year} {2019})}\BibitemShut {NoStop}%
\bibitem [{\citenamefont {Latham}\ \emph {et~al.}(1970)\citenamefont {Latham}, \citenamefont {Ewing}, \citenamefont {Press}, \citenamefont {Sutton}, \citenamefont {Dorman}, \citenamefont {Nakamura}, \citenamefont {Toks{\"o}z}, \citenamefont {Wiggins}, \citenamefont {Derr},\ and\ \citenamefont {Duennebier}}]{Latham1970}%
  \BibitemOpen
  \bibfield  {author} {\bibinfo {author} {\bibfnamefont {G.~V.}\ \bibnamefont {Latham}}, \bibinfo {author} {\bibfnamefont {M.}~\bibnamefont {Ewing}}, \bibinfo {author} {\bibfnamefont {F.}~\bibnamefont {Press}}, \bibinfo {author} {\bibfnamefont {G.}~\bibnamefont {Sutton}}, \bibinfo {author} {\bibfnamefont {J.}~\bibnamefont {Dorman}}, \bibinfo {author} {\bibfnamefont {Y.}~\bibnamefont {Nakamura}}, \bibinfo {author} {\bibfnamefont {N.}~\bibnamefont {Toks{\"o}z}}, \bibinfo {author} {\bibfnamefont {R.}~\bibnamefont {Wiggins}}, \bibinfo {author} {\bibfnamefont {J.}~\bibnamefont {Derr}},\ and\ \bibinfo {author} {\bibfnamefont {F.}~\bibnamefont {Duennebier}},\ }\bibfield  {title} {\bibinfo {title} {Passive seismic experiment},\ }\href {https://doi.org/10.1126/science.167.3918.455} {\bibfield  {journal} {\bibinfo  {journal} {Science}\ }\textbf {\bibinfo {volume} {167}},\ \bibinfo {pages} {455} (\bibinfo {year} {1970})}\BibitemShut {NoStop}%
\bibitem [{\citenamefont {Nakamura}(1983)}]{Nakamura1983}%
  \BibitemOpen
  \bibfield  {author} {\bibinfo {author} {\bibfnamefont {Y.}~\bibnamefont {Nakamura}},\ }\bibfield  {title} {\bibinfo {title} {Seismic velocity structure of the lunar mantle},\ }\href {https://doi.org/10.1029/JB088iB01p00677} {\bibfield  {journal} {\bibinfo  {journal} {Journal of Geophysical Research: Solid Earth}\ }\textbf {\bibinfo {volume} {88}},\ \bibinfo {pages} {677} (\bibinfo {year} {1983})}\BibitemShut {NoStop}%
\bibitem [{\citenamefont {{Panning}}\ \emph {et~al.}(2025)\citenamefont {{Panning}}, \citenamefont {{Lognonn{\'e}}}, \citenamefont {{Creighton}}, \citenamefont {{Trippe}}, \citenamefont {{Quetschke}}, \citenamefont {{Majstrorovi{\'c}}},\ and\ \citenamefont {{Jani}}}]{2025arXiv250915452P}%
  \BibitemOpen
  \bibfield  {author} {\bibinfo {author} {\bibfnamefont {M.~P.}\ \bibnamefont {{Panning}}}, \bibinfo {author} {\bibfnamefont {P.}~\bibnamefont {{Lognonn{\'e}}}}, \bibinfo {author} {\bibfnamefont {T.}~\bibnamefont {{Creighton}}}, \bibinfo {author} {\bibfnamefont {J.}~\bibnamefont {{Trippe}}}, \bibinfo {author} {\bibfnamefont {V.}~\bibnamefont {{Quetschke}}}, \bibinfo {author} {\bibfnamefont {J.}~\bibnamefont {{Majstrorovi{\'c}}}},\ and\ \bibinfo {author} {\bibfnamefont {K.}~\bibnamefont {{Jani}}},\ }\bibfield  {title} {\bibinfo {title} {{Potential for Lunar Interior Science by the Gravitational-Wave Detector LILA}},\ }\href {https://doi.org/10.48550/arXiv.2509.15452} {\bibfield  {journal} {\bibinfo  {journal} {arXiv e-prints}\ ,\ \bibinfo {eid} {arXiv:2509.15452}} (\bibinfo {year} {2025})},\ \Eprint {https://arxiv.org/abs/2509.15452} {arXiv:2509.15452 [physics.geo-ph]} \BibitemShut {NoStop}%
  \bibitem [{\citenamefont {{Komatitsch}}\ and\ \citenamefont {{Tromp}}(2002)}]{komatitsch2002spectral}%
  \BibitemOpen
  \bibfield  {author} {\bibinfo {author} {\bibfnamefont {D.}~\bibnamefont {{Komatitsch}}}\ and\ \bibinfo {author} {\bibfnamefont {J.}~\bibnamefont {{Tromp}}},\ }\bibfield  {title} {\bibinfo {title} {Spectral-element simulations of global seismic wave propagation—i. validation},\ }\href@noop {} {\bibfield  {journal} {\bibinfo  {journal} {Geophysical Journal International}\ }\textbf {\bibinfo {volume} {149}},\ \bibinfo {pages} {390} (\bibinfo {year} {2002})}\BibitemShut {NoStop}%
\bibitem [{\citenamefont {{Komatitsch}}\ \emph {et~al.}(2002)\citenamefont {{Komatitsch}}, \citenamefont {{Ritsema}},\ and\ \citenamefont {{Tromp}}}]{komatitsch2002Thespectral}%
  \BibitemOpen
  \bibfield  {author} {\bibinfo {author} {\bibfnamefont {D.}~\bibnamefont {{Komatitsch}}}, \bibinfo {author} {\bibfnamefont {J.}~\bibnamefont {{Ritsema}}},\ and\ \bibinfo {author} {\bibfnamefont {J.}~\bibnamefont {{Tromp}}},\ }\bibfield  {title} {\bibinfo {title} {The spectral-element method, beowulf computing, and global seismology},\ }\href@noop {} {\bibfield  {journal} {\bibinfo  {journal} {Science}\ }\textbf {\bibinfo {volume} {298}},\ \bibinfo {pages} {1737} (\bibinfo {year} {2002})}\BibitemShut {NoStop}%
\end{thebibliography}
\end{document}